\documentclass[%
 reprint,
superscriptaddress,
 amsmath,amssymb,
 aps,
]{revtex4-2}

\usepackage{graphicx}% Include figure files
\usepackage{dcolumn}% Align table columns on decimal point
\usepackage{bm}% bold math
\usepackage[colorlinks,linkcolor=blue, urlcolor=blue, anchorcolor=blue, citecolor=blue]{hyperref}% add hypertext capabilities
\begin{document}

\preprint{APS/123-QED}

\title{Extreme expected values and their applications in quantum information processing}% Force line breaks with \\
\author{Wangjun Lu}
\email{wjlu1227@zju.edu.cn}
\affiliation{Zhejiang Institute of Modern Physics, Department of Physics, Zhejiang University, Hangzhou 310027, China}
\affiliation{Department of Maths and Physics, Hunan Institute of Engineering, Xiangtan 411104, China}
%\altaffiliation{Physics Department, XYZ University.}%Lines break automatically or can be forced with \\
\author{Lei Shao}
\author{Xingyu Zhang}
\author{Zhucheng Zhang}
\author{Jie Chen}
\affiliation{Zhejiang Institute of Modern Physics, Department of Physics, Zhejiang University, Hangzhou 310027, China}
\author{Hong Tao}
\affiliation{MOE Key Laboratory of Fundamental Physical Quantities Measurement and Hubei Key Laboratory of Gravitation and Quantum Physics, PGMF and School of Physics, Huazhong University of Science and Technology, Wuhan 430074,  China}

%\author{Le-Man Kuang$^{3}$}
\author{Xiaoguang Wang}%
\email{xgwang1208@zju.edu.cn}

\affiliation{Zhejiang Institute of Modern Physics, Department of Physics, Zhejiang University, Hangzhou 310027, China}
%$^{3}$Key Laboratory of Low-Dimensional Quantum Structures and Quantum Control of Ministry of Education,
% 	Department of Physics and Synergetic Innovation Center for Quantum Effects and Applications,
% 	Hunan Normal University, Changsha 410081, China}%Lines break automatically or can be forced with \\


\date{\today}% It is always \today, today,

\begin{abstract}
We consider the probability distribution when the monotonic function $F(X)$ of the independent variable $X$ takes the maximum or minimum expected value under the two constraints of a certain probability and a certain expected value of the independent variable $X$. We proposed an equal probability and equal expected value splitting method. With this method, we proved four inequalities, and two of them can be reduced to Jensen's inequalities. Subsequently, we find that after dividing the non-monotone function $H(X)$ into multiple monotone intervals, the problem of solving the maximum and minimum expected values of $H(X)$ can be transformed into the problem of solving the extreme value of a multiple-variable function. Finally, we apply the proved theory to solve three problems in quantum information processing. When studying the quantum parameter estimation in Mach-Zehnder interferometer, for an equal total input photon number, we find an optimal path-symmetric input state that makes the quantum Fisher information take the maximum value, and we prove that the NOON state is the path-symmetric state that makes the quantum Fisher information takes the minimum value. When studying the quantum parameter estimation in Landau-Zener-Jaynes-Cummings model, we find the optimal initial state of the cavity field that makes the system obtain the maximum quantum Fisher information. Finally, for an equal initial average photon number, we find the optimal initial state of the cavity field that makes the Tavis-Cummings quantum battery have the maximum stored energy and the maximum average charging power.
\end{abstract}

\maketitle

%\tableofcontents

\section{introduction}
In almost all fields, the question of how to find the maximum or minimum expected value of a quantity in a system is very important. For example, in reinforcement learning, the reward of the future guides the direction of the decision. Because of the inevitable uncertainty of future information, it is generally replaced by the expected value of the reward. Therefore, the expected value plays a very important role in strategy improvement in reinforcement learning \cite{russell2010artificial,lee2012neural,sutton2018reinforcement}. In quantum mechanics, almost all problems end up with the average of operators over some quantum state. As a simple example, the average value of the operator $\hat{A}$ in a quantum state $\hat{\rho}$ is $\langle \hat{A}\rangle=\text{Tr}(\hat{\rho}\hat{A})$, where '$\text{Tr}$' denotes trace operation \cite{weinberg1995quantum,holevo2003statistical}. 
The quantum state $\hat{\rho}$ can be expressed in terms of $N$ pure state
\begin{equation}
\hat{\rho}=\sum_{k=1}^{N}p_{k}|\phi_{k}\rangle\langle\phi_{k}|,  \label{1}   
\end{equation}
where $\{|\phi_{k}\rangle\}$ is some set of pure states, not necessarily orthogonal, and $p_{k}$ is the weight of corresponding pure state $|\phi_{k}\rangle$. The trace of $\hat{\rho}^{2}$ is equal to one when the quantum state $\hat{\rho}$ is a pure state, and less than one when $\hat{\rho}$ is a mixed state. Another criterion is that the von Neumann entropy of $\hat{\rho}$ is zero for a pure state, and positive for a mixed state \cite{holevo2003statistical}. The average value of the operator $\hat{A}$ is
\begin{equation}
\langle \hat{A}\rangle=\sum_{k=1}^{N}p_{k}\langle\phi_{k}|\hat{A}|\phi_{k}\rangle,  \label{2}
\end{equation}
which is the expectation value of $\langle\phi_{k}|\hat{A}|\phi_{k}\rangle$ with probability distribution $p_{k}$. If $N=1$,  $\hat{\rho}$ is a pure state, then the expectation value of $\hat{A}$ is
\begin{eqnarray}
\langle \hat{A}\rangle
=\sum_{i=1}^{M}p_{i}A_{i},
\label{3}
\end{eqnarray}
where $M$ is the Hilbert space dimension of $\hat{A}$,  $p_{i}=|\langle \varphi_{i}|\phi_{1}\rangle|^{2}$ and $|\varphi_{i}\rangle$  are the eigenstates of $\hat{A}$. $A_{i}$ is the eigenvalue of operator $\hat{A}$ corresponding to the eigenstate $|\varphi_{i}\rangle$. Obviously, the average value of operator $\hat{A}$ is actually the expectation value of the eigenvalue of the operator $\hat{A}$ with the probability distribution $p_{i}$. In quantum mechanics, we are actually ultimately concerned with solving for the probability distribution when these expectations take their maximum or minimum values, that is, finding the optimal state.

Due to the finiteness of resources, it is necessary to find the optimal  corresponding to the maximum or minimum expectation value of the operator under the same certain constraints, such as the same number of input particles in quantum physics. The Lagrange multiplier method is the most common method for solving conditional extreme value problems. The basic idea of the Lagrange multiplier method is to transform an optimization problem with $n$ variables and $m$ constraints into an extreme value problem with a system of equations with $n + m$ variables \cite{rockafellar1993lagrange}. For example, under the normalization condition and the mean value constraint of the independent variable $X$ ($X\in[0, +\infty)$), i.e., $\int_{0}^{+\infty}p(X)dX=1$ and $\int_{0}^{+\infty}Xp(X)dX=\mu$ . It is easy to exploit the Lagrange multiplier method to find that the probability distribution that makes the entropy $S=-\int_{0}^{+\infty}p(X)\ln p(X)dX$ takes the maximum value is $p(X)=\frac{1}{\mu}\exp(-\frac{X}{\mu})$ \cite{guiasu1985principle}. And the probability distribution when the entropy $S$ takes the maximum value is different under different constraints. However, under the normalization condition as well as the mean value constraint, if a function $F(X)$ does not depend on the probability distribution $p(X)$ and is a more complicated function, the Lagrange multiplier method encounters difficulties in finding the probability distribution that makes the function $F(X)$ have the maximum or minimum expected value.  

In this paper, based on this problem above, we propose the equal probability and equal expected value splitting method. By this method, we first find the probability distribution $p(X)$ that makes the function $F(X)$ take the maximum or minimum expected values when the average slope of $F(X)$ is monotonic. Second, for an arbitrary function $H(X)$, we divide the average slope of $H(X)$ into monotonically increasing and monotonically decreasing multiple intervals so that we can transform the problem of finding the probability distribution when $H(X)$ takes the maximum or minimum expected value into the problem of solving the extreme value of a multivariate function. Finally, using our proved optimal probability distribution and our proposed equal probability and equal expected value splitting method, we find optimal and worst path-symmetric entangled states in the Mach-Zehnder interferometer. By our maximum and minimum expected value theorems, we find the optimal initial state of the optical field when the quantum Fisher information of the system takes the maximum value in the Landau-Zener-Jaynes-Cummings model. Also, we find optimal initial states in the Tavis-Cummings quantum battery.

\section{Two  theorems on extreme expected value}
Let $X$ be a random variable with a series number $x_{1},x_{2},\ldots ,x_{k}$ occurring with probabilities $ p(x_{1}), p(x_{2}),\ldots , p(x_{k})$, respectively. Without loss of generality, we assume $x_{1}<x_{2}<\ldots <x_{k}$. There are two constraints
\begin{eqnarray}
\sum_{i=1}^{k}p(x_{i})&=&\bar{p},   \label{4}\\
\sum_{i=1}^{k}x_{i}p(x_{i})&=&\bar{n}, \label{5}
\end{eqnarray}
where $\bar{p}$ is the sum of all probabilities and $\bar{p}\in(0,1]$, $\bar{n}$ is the expected value of $X$. There are three probability distributions, $p(X)$, $p_{1}(X)$, and $p_{2}(X)$. $p(X)$ is an arbitrary probability distribution. $p_{1}(X)$ and $p_{2}(X)$ are two special probability distributions. The three probability distributions satisfy the two constraints above. The two specific probability distributions $p_{1}(X)$ and $p_{2}(X)$ are
\begin{eqnarray}
p_{1}(X)&=&\frac{x_{m+1}\bar{p}-\bar{n}}{x_{m+1}-x_{m}}\delta_{X,x_{m}}+\frac{\bar{n}-x_{m}\bar{p}}{x_{m+1}-x_{m}}\delta_{X,x_{m+1}},    \label{6}\\
p_{2}(X)&=&\frac{x_{k}\bar{p}-\bar{n}}{x_{k}-x_{1}}\delta_{X,x_{1}}+\frac{\bar{n}-x_{1}\bar{p}}{x_{k}-x_{1}}\delta_{X,x_{k}} , \label{7}
\end{eqnarray}
where $x_{m}$ and $x_{m+1}$ are determind by the inequalities $x_{m}\leq \bar{n}/\bar{p}<x_{m+1}$, and  $x_{1}\leq x_{m}\leq x_{k}$.

In this paper, we proved two theorems, which are present in the following. 

\textbf{Theorem 1}. If the average slope of a function $F(X)$ is getting smaller in the interval $[x_{1},x_{k}]$, then
\begin{eqnarray}
&&\frac{x_{m+1}\bar{p}-\bar{n}}{x_{m+1}-x_{m}}F(x_{m})+\frac{\bar{n}-x_{m}\bar{p}}{x_{m+1}-x_{m}}F(x_{m+1})\geq\nonumber \\
&&\sum_{i=1}^{k}p(x_{i})F(x_{i})\geq \frac{x_{k}\bar{p}-\bar{n}}{x_{k}-x_{1}}F(x_{1}) +\frac{\bar{n}-x_{1}\bar{p}}{x_{k}-x_{1}}F(x_{k}).  \label{8}
\end{eqnarray}

\textbf{Theorem 2}. If the average slope of the function $F(X)$ is getting larger in the interval $[x_{1},x_{k}]$, then
\begin{eqnarray}
&&\frac{x_{m+1}\bar{p}-\bar{n}}{x_{m+1}-x_{m}}F(x_{m})+\frac{\bar{n}-x_{m}\bar{p}}{x_{m+1}-x_{m}}F(x_{m+1})\leq\nonumber \\
&&\sum_{i=1}^{k}p(x_{i})F(x_{i})\leq \frac{x_{k}\bar{p}-\bar{n}}{x_{k}-x_{1}}F(x_{1}) +\frac{\bar{n}-x_{1}\bar{p}}{x_{k}-x_{1}}F(x_{k}).  \label{9}
\end{eqnarray}
 
When $\bar{p}=1$ and $x_{m}=\bar{n}$, we can get the following Jensen's inequalities \cite{jensen1906fonctions} from the above inequalities 
\begin{eqnarray}
&&\sum_{i=1}^{k}p(x_{i})F(x_{i})\leq F(\bar{n}), F(X)\hspace{0.1cm} \text{is concave},   \label{10}\\
&&\sum_{i=1}^{k}p(x_{i})F(x_{i})\geq F(\bar{n}), F(X)\hspace{0.1cm} \text{is  convex}.   \label{11}
\end{eqnarray}
It is worth noting that the increasing or decreasing average slope of a function is equivalent to the description of the function as a convex or concave function. In the following, we will prove the above two theorems.

First, we consider two probability distributions $p(X)$ and $p_{1}(X)$ of the random variable $X$. To facilitate the understanding of the equal probability and equal expected value splitting method, we consider $x_{m}=\bar{n}/\bar{p}$, and the complete  proof of the left inequalities in the two theorems are shown in Appendix \ref{Appendix A} when $x_{m}<\bar{n}/\bar{p}<x_{m+1}$. The two probability distributions satisfy the following relationship
\begin{equation}
\sum_{X=x_{1}}^{x_{k}}p(X)=\sum_{X=x_{1}}^{x_{k}}p_{1}(X)=\bar{p},  \label{12}
\end{equation}
The random variable $X$ has an equal expected value under these two probability distributions
\begin{equation}
\sum_{X=x_{1}}^{x_{k}}Xp(X)=\sum_{X=x_{1}}^{x_{k}}Xp_{1}(X)=x_{m} \bar{p}=\bar{n}.  \label{13}
\end{equation}
Then $x_{m}$ can be determined by
\begin{equation}
x_{m}=\frac{\bar{n}}{\bar{p}}. \label{14}
\end{equation}

We consider a function $F(X)$ of random variable $X$. The difference between the expected value of $F(X)$ under the two probability distributions is
\begin{eqnarray}
\varDelta F&=&\sum_{X=x_{1}}^{x_{k}}p_{1}(X)F(X)-\sum_{X=x_{1}}^{x_{k}}p(X)F(X) \nonumber\\
&=&\bar{p}F(x_{m})-\sum_{X=x_{1}}^{x_{k}}p(X)F(X). \label{15} 
\end{eqnarray}
Substituting Eq.~(\ref{12}) into the above equation yields
\begin{equation}
\varDelta F=\sum_{X=x_{1}}^{x_{k}}p(X)[F(x_{m})-F(X)].  \label{16}
\end{equation}
We divide the probability distribution $p(X)$ into three regions
\begin{eqnarray}
\sum_{X=x_{1}}^{x_{k}}p(X)&=&
\sum_{X=x_{1}}^{x_{m-1}}p(X)+p(x_{m})+
\sum_{X=x_{m+1}}^{x_{m+d}}p(X) \nonumber \\
&=&\sum_{i=1}^{m-1}p(x_{i})+p(x_{m})+
\sum_{j=1}^{d}p(x_{m+j}),    \label{17}
\end{eqnarray}
where $d=k-m$. Then the Eq.~(\ref{16}) can be written as
\begin{eqnarray}
\varDelta F&=&\sum_{i=1}^{m-1}p(x_{i})[F(x_{m})-F(x_{i})]\nonumber \\
&&+\sum_{j=1}^{d}p(x_{m+j})[F(x_{m})-F(x_{m+j})].  \label{18}
\end{eqnarray}
So far we have not used the Eq.~(\ref{13}). Substituting Eq.~(\ref{12}) into Eq.~(\ref{13}) yields
\begin{eqnarray}
&&\sum_{i=1}^{m-1}x_{i}p(x_{i})+x_{m}p(x_{m})+
\sum_{j=1}^{d}x_{m+j}p(x_{m+j}) \nonumber \\
=&&x_{m}\sum_{i=1}^{m-1}p(x_{i})+x_{m}p(x_{m})+
x_{m}\sum_{j=1}^{d}p(x_{m+j}). \label{19}
\end{eqnarray}
Then, we can obtain
\begin{equation}
\sum_{j=1}^{d}(x_{m+j}-x_{m})p(x_{m+j})=
\sum_{i=1}^{m-1}(x_{m}-x_{i})p(x_{i}).   \label{20}
\end{equation}

We divide $p(x_{i})$ into $d$ parts, namely, $p(x_{i})=\sum_{j=1}^{d}p_{j}(x_{i})$ where $i=1,2,\cdots,m-1$. Then, Eq.~(\ref{20}) can be written as
\begin{equation}
\sum_{j=1}^{d}\bigg[(x_{m+j}-x_{m})p(x_{m+j})-
\sum_{i=1}^{m-1}(x_{m}-x_{i})p_{j}(x_{i})\bigg]=0.  \label{21}
\end{equation}
If we can divide $p(x_{i})$ into $d$ parts so that all $d$ terms in Eq.~(\ref{21}) are zero, we can obtain the following solution
\begin{equation}
p(x_{m+j})=\frac{\sum_{i=1}^{m-1}(x_{m}-x_{i})p_{j}(x_{i})}{x_{m+j}-x_{m}}.   \label{22}
\end{equation}
In fact, we do find such a splitting method
\begin{equation}
p_{j}(x_{i})=\frac{(x_{m+j}-x_{m})p(x_{m+j})}{\sum_{k=1}^{d}(x_{m+k}-x_{m})p(x_{m+k})}p(x_{i}). \label{23}
\end{equation}
This splitting method is called the equal probability and equal expected value splitting method. For the above equation, it is easy to verify that $\sum_{j=1}^{d}p_{j}(x_{i})=p(x_{i})$ and the above equation can make each term in Eq.~(\ref{21}) equal to zero. Similarly, we divide the probability distribution of $p_{1}(X)$ into the following $d+1$ parts
\begin{eqnarray}
\bar{p}_{j}&=&\frac{(x_{m+j}-x_{m})p(x_{m+j})}{\sum_{k=1}^{d}(x_{m+k}-x_{m})p(x_{m+k})}\sum_{i=1}^{m-1}p(x_{i}) \nonumber \\
&&+p(x_{m+j}) \label{24}\\
\bar{p}_{d+1}&=&p(x_{m}),  \label{25}
\end{eqnarray}
where $j=1, 2, \cdots, d$. From Eq.~(\ref{23}) and Eq.~(\ref{24}), we can obtain the following two equations
\begin{eqnarray}
&&\sum_{i=1}^{m-1}p_{j}(x_{i})+p(x_{m+j})=\bar{p}_{j},  \label{26}\\
&&\sum_{i=1}^{m-1}x_{i}p_{j}(x_{i})+x_{m+j}p(x_{m+j})=x_{m}\bar{p}_{j}. \label{27}
\end{eqnarray}
The left and right sides of Eq.~(\ref{26}) are the $j$-th probability of the probability distributions $p(X)$ and $p_{1}(X)$, respectively. The left and right sides of Eq.~(\ref{27}) are the expected values of the independent variable $X$ under the $j$-th probability distribution of the two probabilities. This is the reason that we call the splitting method of Eq.~(\ref{23}) the equal probability and equal expected value splitting method.

Substituting Eq.~(\ref{22}) into Eq.~(\ref{18}) yields
\begin{eqnarray}
\varDelta F
&=&\sum_{i=1}^{m-1}\sum_{j=1}^{d}p^{j}(x_{i})(x_{m}-x_{i})\bigg[\frac{F(x_{m})-F(x_{i})}{x_{m}-x_{i}} \nonumber \\
&&-\frac{F(x_{m+j})-F(x_{m})}{x_{m+j}-x_{m}}\bigg]. \label{28}
\end{eqnarray}
Since $p^{j}(x_{i})\geq0$ and $(x_{m}-x_{i})>0$, if the average slope of the function $F(X)$ increases with the increase of $X$, then $\frac{F(x_{m})-F(x_{i})}{x_{m}-x_{i}}<\frac{F(x_{m+j})-F(x_{m})}{x_{m+j}-x_{m}}$ and $\varDelta F\leq0$, $\bar{p}F(x_{m})\leq\sum_{X=x_{1}}^{x_{k}}p(X)F(X)$. Conversely, $\frac{F(x_{m})-F(x_{i})}{x_{m}-x_{i}}>\frac{F(x_{m+j})-F(x_{m})}{x_{m+j}-x_{m}}$ and $\varDelta F\geq0$, $\bar{p}F(x_{m})\geq\sum_{X=x_{1}}^{x_{k}}p(X)F(X)$.

From the above and Appendix \ref{Appendix A}, we proved that with the equal sum of probabilities and the same expected value of $X$, if the average slope of the function $F(X)$ decreases as $X$ increases, the probability distribution that makes the function $F(X)$ has the maximum expected value is $p_{1}(X)$. Conversely, if the average slope of the function $F(X)$ increases as $X$ increases, then $p_{1}(X)$ is the probability distribution that makes the function $F(X)$ has the minimum expected value. Assuming that all variables of $X$ are integers, $\bar{p}=1$, and  the expected value $\bar{n}$ of $X$ is a non-integer. In that case, we can split the probability distribution into two parts, with the expected value of $X$ under one part of the probability distribution being an integer and the expected value under the other part of the probability distribution is an additional fractional part. Using the above method, we can show that $p_{1}(X)=[1-(\bar{n}-[\bar{n}])]\delta_{X,[\bar{n}]}+(\bar{n}-[\bar{n}])\delta_{X,[\bar{n}]+1}$ is the probability distribution that makes the function $F(X)$ to take the maximum and minimum expected values when the average slope of the function $F(X)$ increases and decreases, respectively. $[\bar{n}]$ represents the integer part of $\bar{n}$. This specific proof process is described in Ref.~\cite{PhysRevA.104.043706}.

Second, we consider two probability distributions $p(X)$ and $p_{2}(X)=p_{2}(x_{1})\delta_{X,x_{1}}+p_{2}(x_{k})\delta_{X,x_{k}}$ of the random variable $X$. These two probability distributions have an equal sum of probabilities and the same expected value
\begin{eqnarray}
\sum_{i=1}^{k}p(x_{i})&=&\sum_{i=1}^{k}p_{2}(x_{i})=p_{2}(x_{1})+p_{2}(x_{k})=\bar{p},   \label{29} \\
\sum_{i=1}^{k}x_{i}p(x_{i})&=&\sum_{i=1}^{k}x_{i}p_{2}(x_{i})=x_{1}p_{2}(x_{1})+x_{k}p_{2}(x_{k}) \nonumber \\
&=&\bar{n},  \label{30}
\end{eqnarray}
where $\bar{p}\in(0,1]$.

The difference between the expected value of the function $F(X)$ under the two probability distributions $p(X)$ and $p_{2}(X)$ is the following expression
\begin{eqnarray}
\varDelta F&=&\sum_{X=x_{1}}^{x_{k}}p_{2}(X)F(X)-\sum_{X=x_{1}}^{x_{k}}p(X)F(X) \nonumber \\
&=&p_{2}(x_{1})F(x_{1})+p_{2}(x_{k})F(x_{k})-\sum_{i=1}^{k}p(x_{i})F(x_{i}).  \nonumber \\ \label{31}
\end{eqnarray}
From equations Eq.~(\ref{29}) and Eq.~(\ref{30}), we can obtain the solutions for $p_{2}(x_{1})$ and $p_{2}(x_{k})$
\begin{eqnarray}
p_{2}(x_{1})&=&\frac{x_{k}\bar{p}-\bar{n}}{x_{k}-x_{1}}, \label{32}\\
p_{2}(x_{k})&=&\frac{\bar{n}-x_{1}\bar{p}}{x_{k}-x_{1}}.  \label{33}
\end{eqnarray}
Substituting Eq.~(\ref{32}) and Eq.~(\ref{33}) into Eq.~(\ref{31}) gives
\begin{eqnarray}
\varDelta F
&=&\sum_{i=2}^{k-1}\frac{p(x_{i})(x_{i}-x_{1})(x_{k}-x_{i})}{x_{k}-x_{1}}\Bigg[\frac{F(x_{k})-F(x_{i})}{x_{k}-x_{i}} \nonumber  \\
&&-\frac{F(x_{i})-F(x_{1})}{x_{i}-x_{1}}\Bigg].  \label{34}
\end{eqnarray}
The details
of the derivation of the above equation are shown in Appendix \ref{Appendix B}. Since $p(x_{i})\geq0$ and $x_{k}>x_{i}>x_{1}(i=2,3,\cdots,k-1)$, if the average slope of the function $F(X)$ increases with the increase of $X$, then $\frac{F(x_{k})-F(x_{i})}{x_{k}-x_{i}}>\frac{F(x_{i})-F(x_{1})}{x_{i}-x_{1}}$ and $\varDelta F\geq0$, $\sum_{X=x_{1}}^{x_{k}}p_{2}(X)F(X)\geq\sum_{X=x_{1}}^{x_{k}}p(X)F(X)$. Conversely, $\frac{F(x_{k})-F(x_{i})}{x_{k}-x_{i}}<\frac{F(x_{i})-F(x_{1})}{x_{i}-x_{1}}$ and $\varDelta F\leq0$, $\sum_{X=x_{1}}^{x_{k}}p_{2}(X)F(X)\leq\sum_{X=x_{1}}^{x_{k}}p(X)F(X)$.
Namely, $p_{2}(X)$ is the probability distribution that makes the function $F(X)$ has the maximum and minimum expected values when the average slope of the function $F(X)$ increases and decreases, respectively.

In summary, if the average slope of the function $F(X)$ increases as $X$ increases, the probability distributions that make the function $F(X)$ has the maximum and minimum expected values are $p_{2}(X)$ and $p_{1}(X)$, respectively. If the average slope of the function $F(X)$ decreases as $X$ increases, $p_{1}(X)$ and $p_{2}(X)$ are the probability distributions that makes the function $F(X)$ take the maximum and minimum expected values, respectively.

\section{Extreme expected values of an arbitrary function}

In the following, we will show that solving the maximum and minimum expected values of an arbitrary function can be translated into solving the extreme value problem of the function. We solve for the maximum and minimum expectation of the function $H(X)$ under the two conditions of probability normalization and the expected value of $X$ being $\bar{N}$. According to the monotonicity of the average slope of $H(X)$ changing with $X$, we divide the value range of $X$ into multiple intervals $[X_{i}, X_{i+1}]$ ($i=1, 2, \cdots, k-1$), where $X_{1}$ and $X_{k}$ are the end point values, and the other inflection points $X_{i}$ can be obtained by $H''(X)=0$ or $H(X)$ is continuous but not derivable at these points. For ease of description, it is assumed without loss of generality that the average slope of the function $H(X)$ is monotonically decreasing in the first interval and the monotonicity is different in adjacent intervals, i.e., the average slope of $H(X)$ is monotonically decreasing and monotonically increasing when $i$ is an odd number and an even number, respectively. We split the normalized probability and the expected value of $X$ into these $k-1$ intervals. The probability and expected value of each interval are $p_{i}$ and $\bar{n}_{i}$ respectively. These probabilities and expected values satisfy the following two equations
\begin{eqnarray}
\sum_{i=1}^{k-1}p_{i}=1, \label{35}\\
\sum_{i=1}^{k-1}\bar{n}_{i}=\bar{N}. \label{36}
\end{eqnarray}

According to our maximum and minimum expectation theorems, the maximum and minimum expected values of $H(X)$ under the two constraints are
\begin{eqnarray}
\overline{H(X)}_{max}&=&\sum_{j=1}^{(k-1)/2}\Bigg[p_{2j-1}H(\frac{\bar{n}_{2j-1}}{p_{2j-1}}) +\frac{X_{2j+1}p_{2j}-\bar{n}_{2j}}{X_{2j+1}-x_{2j}} \nonumber \\
&&\times H(X_{2j})+\frac{\bar{n}_{2j}-X_{2j}p_{2j}}{X_{2j+1}-X_{2j}} H(X_{2j+1})   \Bigg], \label{37}\\
\overline{H(X)}_{min}&=&\sum_{j=1}^{(k-1)/2}\Bigg[ \frac{X_{2j}p_{2j-1}-\bar{n}_{2j-1}}{X_{2j}-X_{2j-1}}H(X_{2j-1})   \nonumber \\
&&+\frac{\bar{n}_{2j-1}-X_{2j-1}p_{2j-1}}{X_{2j}-X_{2j-1}}H(X_{2j}) \nonumber \\
&&+p_{2j}H(\frac{\bar{n}_{2j}}{p_{2j}})\Bigg].\label{38}
\end{eqnarray}
Here we assume that $k$ is an odd number and $k\geq 3$. There are some additional constraints here, $X_{2j-1}\leq\frac{\bar{n}_{2j-1}}{p_{2j-1}}\leq X_{2j}\leq\frac{\bar{n}_{2j}}{p_{2j}}\leq X_{2j+1}$ in the last two equations.

It is worth noting that we used the condition $\bar{n}_{i}/p_{i}\in \{x_{1}, x_{2}, \cdots, x_{k}\}$ above. In fact, when we split the probability and the expected value of the variable $X$ in each interval, we first do not know the specific probability value and the expected value of the variable $X$ into each interval, which makes us unable to know $x_{m}$ and $x_{m+1}$ in each interval. So we first assume that $\bar{n}_{i}/p_{i}\in \{x_{1}, x_{2}, \cdots, x_{k}\}$, and finally, determine the form of the probability distribution $p_{1}(X)$ by solving the specific values of $\bar{n}_{i}$ and $p_{i}$ in the optimal probability distribution. If we substitute Eq.~(\ref{35}) and Eq.~(\ref{36}) into Eq.~(\ref{37}) and Eq.~(\ref{38}), then Eq.~(\ref{37}) and Eq.~(\ref{38}) are actually two equations with $2(k-2)$ variables. Therefore, we transform the maximum or minimum expectations of solving the function $H(X)$ into the extreme value of solving the multivariate function $\overline{H(X)}_{max}$ or $\overline{H(X)}_{min}$.

\begin{figure}[t]
\centering
\includegraphics[width=9cm,height=7cm]{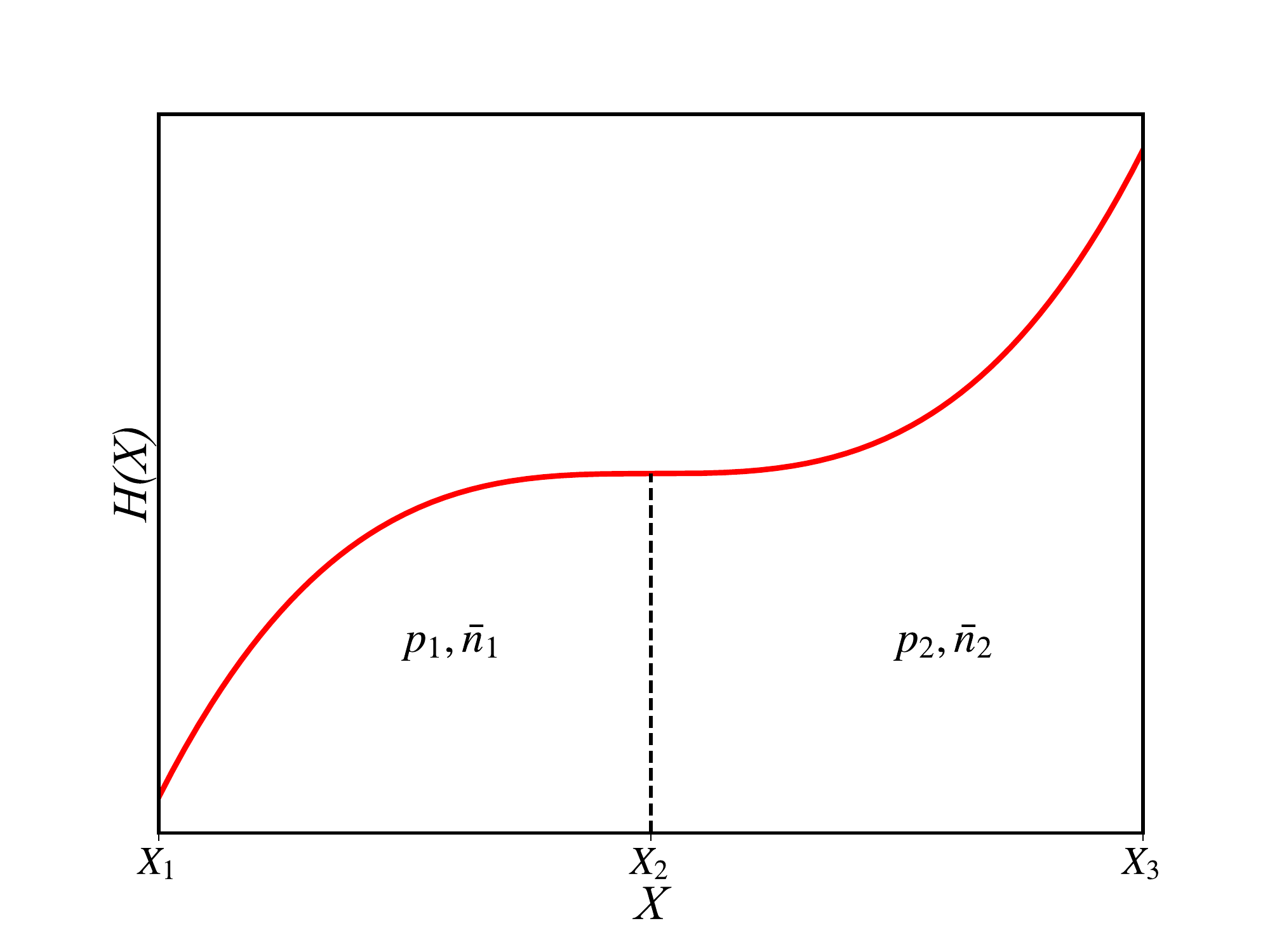}% Here is how to import EPS art
\caption{\label{fig1}Function $H(X)$ varies with the independent variable $X$. The average slope of $H(X)$ is decreasing and increasing in intervals $[X_{1},X_{2}]$ and $[X_{2},X_{3}]$, respectively. $p_{i}$ and $\bar{n}_{i}(i=1,2)$ are the probability and the expected value of the variable $X$, respectively.}
\end{figure}

In the following, we take $k = 3$ as an example. We assume that the image of the function $H(X)$ is shown in Fig.~\ref{fig1}. The average slope of the function $H(X)$ is decreasing in the interval $[X_{1},X_{2}]$, while it is increasing in the interval $[X_{2},X_{3}]$. The inflection point $X_{2}$ can be obtained by $H''(X)=0$. The sums of the probabilities and the expected values of $X$ we assign in these two intervals are $p_{i}$ and $\bar{n}_{i}$ $(i=1,2)$
\begin{eqnarray}
p_{1}+p_{2}&=&1, \label{39} \\
\bar{n}_{1}+\bar{n}_{2}&=&\bar{N}.   \label{40}
\end{eqnarray}
According to our maximum and minimum expectation theorems, the maximum and minimum expected values of $H(X)$ are 
\begin{eqnarray}
\overline{H(X)}_{max}&=&p_{1}H\Bigg((\frac{\bar{n}_{1}}{p_{1}}\Bigg)+\frac{X_{3}p_{2}-\bar{n}_{2}}{X_{3}-X_{2}}H(X_{2}) \nonumber \\
&&+\frac{\bar{n}_{2}-X_{2}p_{2}}{X_{3}-X_{2}}H(X_{3}),     \label{41}\\
\overline{H(X)}_{min}&=&\frac{X_{2}p_{1}-\bar{n}_{1}}{X_{2}-X_{1}}H(X_{1})+\frac{\bar{n}_{1}-X_{1}p_{1}}{X_{2}-X_{1}}H(X_{2}) \nonumber \\
&&+p_{2}H\Bigg(\frac{\bar{n}_{2}}{p_{2}}\Bigg).  \label{42}
\end{eqnarray}
Some additional constraints will be introduced here, i.e., $X_{1}\leq \bar{n}_{1}/p_{1}\leq X_{2}\leq \bar{n}_{2}/p_{2}\leq X_{3}$ in Eq.~(\ref{41}) and Eq.~(\ref{42}). Substituting Eq.~(\ref{39}) and Eq.~(\ref{40}) into Eq.~(\ref{41}) and Eq.~(\ref{42}) yields
\begin{eqnarray}
\overline{H(X)}_{max}&=&(1-p_{2})H(\frac{\bar{N}-\bar{n}_{2}}{1-p_{2}})+\frac{X_{3}p_{2}-\bar{n}_{2}}{X_{3}-X_{2}}H(X_{2}) \nonumber \\
&&+\frac{\bar{n}_{2}-X_{2}p_{2}}{X_{3}-X_{2}}H(X_{3}) ,  \label{43}\\
\overline{H(X)}_{min}&=&\frac{X_{2}(1-p_{2})-(\bar{N}-\bar{n}_{2})}{X_{2}-X_{1}}H(X_{1})+p_{2}H(\frac{\bar{n}_{2}}{p_{2}}) \nonumber \\
&&+\frac{(\bar{N}-\bar{n}_{2})-X_{1}(1-p_{2})}{X_{2}-X_{1}}H(X_{2}).  \label{44}
\end{eqnarray}
$\overline{H(X)}_{max}(p_{2},\bar{n}_{2})$ and $\overline{H(X)}_{min}(p_{2},\bar{n}_{2})$ are both binary functions. Therefore, we transform the maximum or minimum expected value problem of the function into the extreme value problem of the function using the maximum and minimum expected value theorems we proved.

\section{Applications of the Maximum and Minimum Expected value theorems}
In this section, we use the maximum and minimum expected value theorems proved above to study some fundamental problems in quantum information processing. Three examples of our theoretical applications are given here, two of them is to find the initial state when the system has the maximum or minimum quantum Fisher information, and the last is to find the initial state when the system has the maximum stored energy and the maximum average charging power.

\subsection{Quantum parameter estimation in the Mach-Zehnder interferometer}
We consider a Mach-Zehnder (M-Z) interferometer. We aim to find the initial state when the quantum Fisher information of the output state takes the maximum value for the same number of input particles. The input state $|\Psi_\text{in}\rangle$ of the interferometer is as follows \cite{lee2016quantum}
\begin{equation}
|\Psi_{\text{in}}\rangle=C(|\phi\rangle_\text{A}|0\rangle_\text{B}+|0\rangle_\text{A}|\phi\rangle_\text{B}),  \label{45}
\end{equation}
where $C=1/\sqrt{2(1+|\langle0|\phi\rangle|^{2})}$ is the normalization coefficient. $|\Psi_\text{in}\rangle$ is prepared in modes $A$ and $B$. $|\phi\rangle_{i}$ is an arbitrary state in mode $i$ $(i=A, B)$. The input state $|\Psi_\text{in}\rangle$ is called NOON states \cite{PhysRevA.40.2417, PhysRevLett.85.2733,lee2002quantum} and entangled coherent states \cite{PhysRevA.45.6811,sanders2012review,liu2016quantum} when the state $|\phi\rangle_{i}$ $(i=A, B)$ is a number state $|N\rangle$ and a coherent state $|\alpha\rangle$, respectively. Generally, $|\Psi_\text{in}\rangle$ is called a path-symmetric entangled state.

\begin{figure}[t]
\centering
\includegraphics[width=8.3cm,height=6cm]{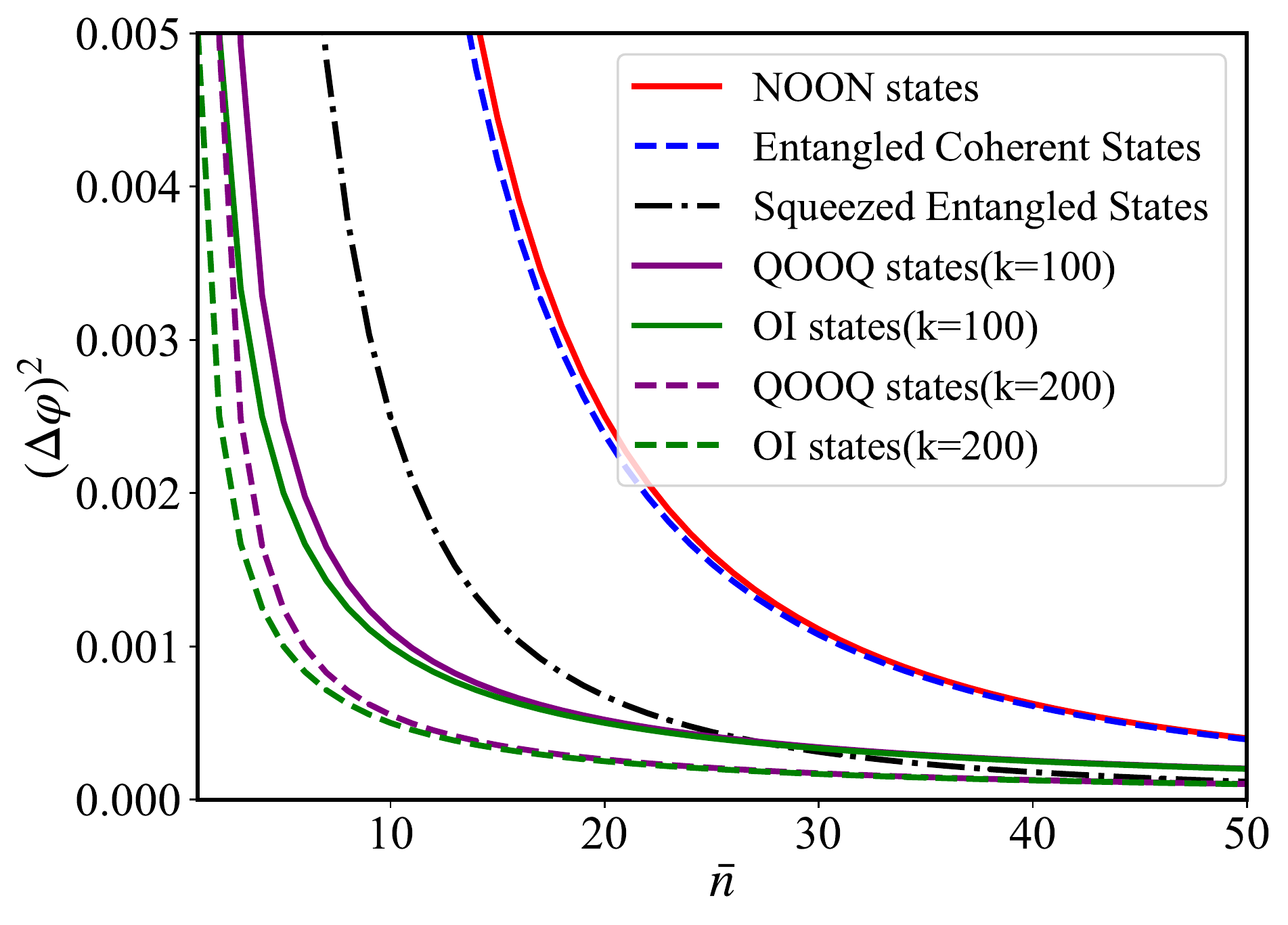}% Here is how to import EPS art
\caption{\label{fig2}Variation of quantum Cramer-Rao bound with the total input average photon number for different  path-symmetric entangled states.}
\end{figure}

The operation of the M-Z interferometer on the input state is a unitary operation. When the phase-shift occurs in both of the arms of the M-Z interferometer, the unitary operation is represented by the following operator
\begin{equation}
\hat{U}(\varphi)=e^{-i\varphi(\hat{n}_\text{A}-\hat{n}_\text{B})/2}, \label{46}
\end{equation}
where $\varphi$ is the relative phase between the two arms of the M-Z interferometer and $\hat{n}_{i}$ is the bosonic number operator in mode $i$ $(i=A, B)$. Then the output state of the M-Z interferometer is as follows
\begin{eqnarray}
|\Psi_\text{out}\rangle&=&\hat{U}(\varphi)|\Psi_\text{in}\rangle \nonumber \\
&&=e^{-i\varphi(\hat{n}_\text{A}-\hat{n}_\text{B})/2}|\Psi_\text{in}\rangle. \label{47}
\end{eqnarray}
Since the output state is a pure state, the quantum Fisher information for the output state is
\begin{equation}
F=4[\langle\partial_{\varphi}\Psi_\text{out}|\partial_{\varphi}\Psi_\text{out}\rangle-|\langle\Psi_\text{out}|\partial_{\varphi}\Psi_\text{out}\rangle|^{2}].  \label{48}
\end{equation}
Substituting Eq.~(\ref{41}) and Eq.~(\ref{43}) into the last equation yields
\begin{equation}
F=\sum_{n=0}^{k}p(n)f(n),  \label{49}
\end{equation}
where $p(n)=|\langle n|\phi\rangle|^{2}$ is the modulus square of the inner product between the number state $|n\rangle$ and an arbitrary state $|\phi\rangle$, $k$ is the maximum value among the variables
$n$ whose corresponding probability $p(k)$ is not zero, and 
\begin{equation}
f(n)=\frac{n^{2}}{1+p(0)},  \label{50}
\end{equation}
where $p(0)=|\langle 0|\phi\rangle|^{2}$. Obviously, the quantum Fisher information $F$ is an expected value of the $f(n)$. Since the average slope of $f(n)$ increases as $n$ increases, the optimal probability distribution $p_\text{O}(n)$ when $f(n)$ takes the maximum expected value according to our maximum and minimum expectation theorem is
\begin{equation}
p_\text{O}(n)=\frac{k-\bar{n}}{k+\bar{n}}\delta_{0,n}+\frac{2\bar{n}}{k+\bar{n}}\delta_{k,n}, \label{51}
\end{equation}
where $\bar{n}$ is the input total photon number, i.e., $\bar{n}=\langle\Psi_\text{in}|(\hat{n}_\text{A}+\hat{n}_\text{B})|\Psi_\text{in}\rangle$. $k$ is the mean value of the number operator $\hat{n}_{i}$ ($i=A, B$) in number state $|k\rangle$, and $k\geq \bar{n}$. The specific value of $k$ we discuss later. It is worth noting that our theorems cannot be simply applied here. The probability distribution in the above equation is obtained by substituting $|\phi\rangle_{i\text{O}}=\sqrt{p(0)}|0\rangle+\sqrt{p(k)}|k\rangle$ ($i=A, B$) into $\bar{n}=\langle\Psi_\text{in}|(\hat{n}_\text{A}+\hat{n}_\text{B})|\Psi_\text{in}\rangle$ for the solution. With an equal input total photon number, we immediately obtain the optimal input state (OI state) that makes the quantum Fisher information has the maximum value as
\begin{equation}
|\Psi_\text{in}\rangle_\text{OI}=C_\text{O}(|\phi\rangle_\text{AO}|0\rangle_\text{B}+|0\rangle_\text{A}|\phi\rangle_\text{BO}), \label{52}
\end{equation}
where 
\begin{eqnarray}
|\phi\rangle_{i\text{O}}&=&\sqrt{\frac{k-\bar{n}}{k+\bar{n}}}|0\rangle+\sqrt{\frac{2\bar{n}}{k+\bar{n}}}|k\rangle, i=\text{A, B}, \label{53} \\
C_\text{O}&=&\sqrt{\frac{k+\bar{n}}{4k}}. \label{54}
\end{eqnarray}
$|\phi\rangle_{i\text{O}}$ is the superposition of the number states $|0\rangle$ and $|k\rangle$. We can verify the above conclusion. The difference between the quantum Fisher information in the optimal input state and an arbitrary input state is
\begin{eqnarray}
\varDelta F_\text{O}=F_\text{O}-F. \label{55}
\end{eqnarray}
$F_\text{O}$ and $F$ are the quantum Fisher information with the input state $|\Psi_\text{in}\rangle$ as the optimal state and an arbitrary state, respectively.
Substituting Eq.~(\ref{47}) and Eq.~(\ref{52}) into Eq.~(\ref{48}), we get
\begin{equation}
F_\text{O}=k\bar{n} . \label{56}
\end{equation}
Since the arbitrary state $|\phi\rangle_{i}$ ($i=A, B$) and the optimal state $|\phi\rangle_{i\text{O}}$ ($i=A, B$) have the same input total photon number, we can obtain the following relationship
\begin{equation}
\bar{n}=\sum_{n=0}^{N}p(n)\frac{n}{1+p(0)}.  \label{57}
\end{equation}
Substituting Eq.~(\ref{49}), Eq.~(\ref{56}), and Eq.~(\ref{57}) into Eq.~(\ref{55}) yields
\begin{equation}
\varDelta F_\text{O}=\sum_{n=0}^{N}p(n)\frac{n}{1+p(0)}(k-n). \label{58}
\end{equation}
According to our theorems, the value of $k$ should be infinite, so $k\geq n$ and $\varDelta F_\text{O}\geq0$, that is, the state $|\Psi_\text{in}\rangle_\text{OI}$ is the input state which makes the quantum Fisher information takes the maximum value. It is worth noting that the quantum Fisher information in the optimal initial state is $F_\text{O}=k\bar{n}$, and the quantum Cramer-Rao bound is $(\varDelta \varphi)^{2}\geqslant 1/F_\text{O}=1/(k\bar{n})$. If $k=\bar{n}$, the input state is a NOON state whose quantum Cramer-Rao bound is $(\varDelta \varphi)^{2}\geqslant 1/(\bar{n})^{2}$, which is the Heisenberg quantum limit. If $k>\bar{n}$, then the quantum Cramer-Rao bound exceeds the Heisenberg quantum limit ($\sim 1/(\bar{n})^{2}$). Theoretically, the value of $k$ can be infinite, and then the quantum Cramer-Rao bound is infinitesimal. That is,  $|\Psi_\text{in}\rangle_\text{OI}$ is such a state which has an infinite quantum Fisher information when the average photon number is finite. This interesting result is also shown in Ref.~\cite{PhysRevLett.62.2377,PhysRevA.43.3795,PhysRevA.44.3365,PhysRevLett.104.103602,rivas2012sub,zhang2012unbounded,PhysRevA.95.032113,PhysRevA.104.012607} and has been discussed extensively \cite{PhysRevLett.108.230401,PhysRevLett.108.210404,PhysRevA.88.060101,PhysRevX.5.031018}.  In Fig.~\ref{fig2}, we compare the quantum Cramer-Rao bound of the NOON state \cite{PhysRevA.54.R4649}, 
Entangled coherent state \cite{PhysRevLett.107.083601}, Squeezed entangled state \cite{lee2016quantum,Shao:20}, QOOQ state \cite{lee2016quantum}, and our OI state ($|\Psi_\text{in}\rangle_\text{OI}$) with the total input average photon number. We found that as long as the value of $k$ is large enough, the state with the smallest quantum Cramer-Rao bound under the equal input total photon number is the OI state.

In the following, we use our theorems to find the path-symmetric entangled state that makes the M-Z interferometer has the minimum quantum Fisher information. According to our theorems, when $\bar{n}$ is an integer, the quantum Fisher information is minimized when $|\phi\rangle_{i}=|\bar{n}\rangle$ ($i=A, B$) for the equal input total photon number $\bar{n}$, i.e., the NOON state $(|\bar{n}\rangle|0\rangle+|0\rangle|\bar{n}\rangle)/\sqrt{2}$ in the path-symmetric entangled state is the input state that makes the quantum Fisher information takes the minimum value. We will demonstrate this briefly below.

The difference in the quantum Fisher information between the NOON state and an arbitrary path-symmetric entangled state is
\begin{eqnarray}
\varDelta F_\text{NOON}&=&\bar{n}^{2}-\sum_{n=0}^{N}p(n)\frac{n^{2}}{1+p(0)} \nonumber \\
&=&\sum_{n=0}^{N}p(n)\Big[\bar{n}^{2}-\frac{n^{2}}{1+p(0)}\Big]. \label{59}
\end{eqnarray}
We have used $\sum_{n=0}^{N}p(n)=1$ above. From the equal input total photon number for the NOON state and an arbitrary path-symmetric entangled state, we can obtain the following relationship
\begin{eqnarray}
\bar{n}&=&\sum_{n=0}^{N}p(n)\frac{n}{1+p(0)}  \nonumber \\
&=&\sum_{n=0}^{N}p(n)\bar{n}. \label{60}
\end{eqnarray}
We divide the probability distribution into two parts
\begin{eqnarray}
&&\sum_{n=0}^{m}p(n)\frac{n}{1+p(0)}+\sum_{j=1}^{d}p(m+j)\frac{m+j}{1+p(0)} \nonumber \\
&&=\sum_{n=0}^{m}p(n)\bar{n}+\sum_{j=1}^{d}p(m+j)\bar{n},  \label{61}
\end{eqnarray}
where $m$ is determined by the inequality $m/(1+p(0))\leq \bar{n}\leq (m+1)/(1+p(0))$, and $d=N-m$.
Then we obtain
\begin{equation}
\sum_{j=1}^{d}p(m+j)\Big[\frac{m+j}{1+p(0)}-\bar{n}\Big]=\sum_{n=0}^{m}p(n)\Big[\bar{n}-\frac{n}{1+p(0)}\Big]. \label{62}
\end{equation}
We divide $p(n)$ into $d$ parts, namely, $p(n)=\sum_{j=1}^{d}p_{j}(n)$, where $n\in[0,m]$.
The expression of $p_{j}(n)$ is as follows
\begin{equation}
p_{j}(n)=\frac{p(m+j)\Big[\frac{m+j}{1+p(0)}-\bar{n}\Big]}{\sum_{k=1}^{d}p(m+k)\Big[\frac{m+k}{1+p(0)}-\bar{n}\Big]}p(n).  \label{63}
\end{equation}
Then Eq.~(\ref{62}) can be written as
\begin{eqnarray}
&&\sum_{j=1}^{d}\Bigg[p(m+j)\Big[\frac{m+j}{1+p(0)}-\bar{n}\Big]-\sum_{n=0}^{m}p^{j}(n)\Big[\bar{n}-\frac{n}{1+p(0)}\Big]\Bigg] \nonumber \\
&&=0  . \label{64}
\end{eqnarray}
For $p_{j}(n)$ of Eq.~(\ref{63}), each term in the above formula is equal to zero. Then we can obtain
\begin{equation}
p(m+j)=\frac{\sum_{n=0}^{m}p^{j}(n)\Big[\bar{n}-\frac{n}{1+p(0)}\Big]}{\frac{m+j}{1+p(0)}-\bar{n}} .\label{65}
\end{equation}
Substituting Eq.~(\ref{65}) into Eq.~(\ref{59}) yields
\begin{eqnarray}
&&\varDelta F_\text{NOON}
=\sum_{n=0}^{m}\sum_{j=1}^{d}p^{j}(n)\Big[\bar{n}-\frac{n}{1+p(0)}\Big]\times \nonumber \\
&&\Bigg[\frac{(1+p(0))\bar{n}^{2}-n^{2}}{(1+p(0))\bar{n}-n}-\frac{(m+j)^{2}-\bar{n}^{2}(1+p(0))}{m+j-\bar{n}(1+p(0))}\Bigg]. \label{66}
\end{eqnarray}
The specific calculation of the above equation is shown in Appendix \ref{Appendix C}. It is easy to prove the following two inequalities hold
\begin{eqnarray}
\frac{(1+p(0))\bar{n}^{2}-n^{2}}{(1+p(0))\bar{n}-n}&\leqslant& \frac{\bar{n}^{2}-n^{2}}{\bar{n}-n}, \label{67} \\
\frac{(m+j)^{2}-\bar{n}^{2}}{m+j-\bar{n}}&\leqslant&\frac{(m+j)^{2}-\bar{n}^{2}(1+p(0))}{m+j-\bar{n}(1+p(0))}. \label{68}
\end{eqnarray}
Since the average slope of the function $g(n)=n^{2}$ is increasing, $\frac{\bar{n}^{2}-n^{2}}{\bar{n}-n}\leqslant\frac{(m+j)^{2}-\bar{n}^{2}}{m+j-\bar{n}}$. Then
\begin{eqnarray}
\frac{(1+p(0))\bar{n}^{2}-n^{2}}{(1+p(0))\bar{n}-n}\leqslant\frac{(m+j)^{2}-\bar{n}^{2}(1+p(0))}{m+j-\bar{n}(1+p(0))}.\label{69}
\end{eqnarray}
And since $p^{j}(n)\geqslant 0$ and $\bar{n}-\frac{n}{1+p(0)}\geqslant 0$, $\varDelta F_\text{NOON}\leqslant 0$, i.e., the NOON state in the path-symmetric entangled states is the input state  that makes the quantum Fisher information takes the minimum value. This is the reason why any other state is better than the NOON state in Fig.~\ref{fig2} and  Refs.~\cite{Shao:20,PhysRevLett.107.083601,lee2016quantum}, and all path-symmetric pure states can achieve their maximal phase sensitivity \cite{PhysRevA.79.033822}.

\subsection{Quantum parameter estimation in the Landau-Zener-Jaynes-Cummings model}

\begin{figure}[h!]
\centering
\includegraphics[width=8.8cm,height=6cm]{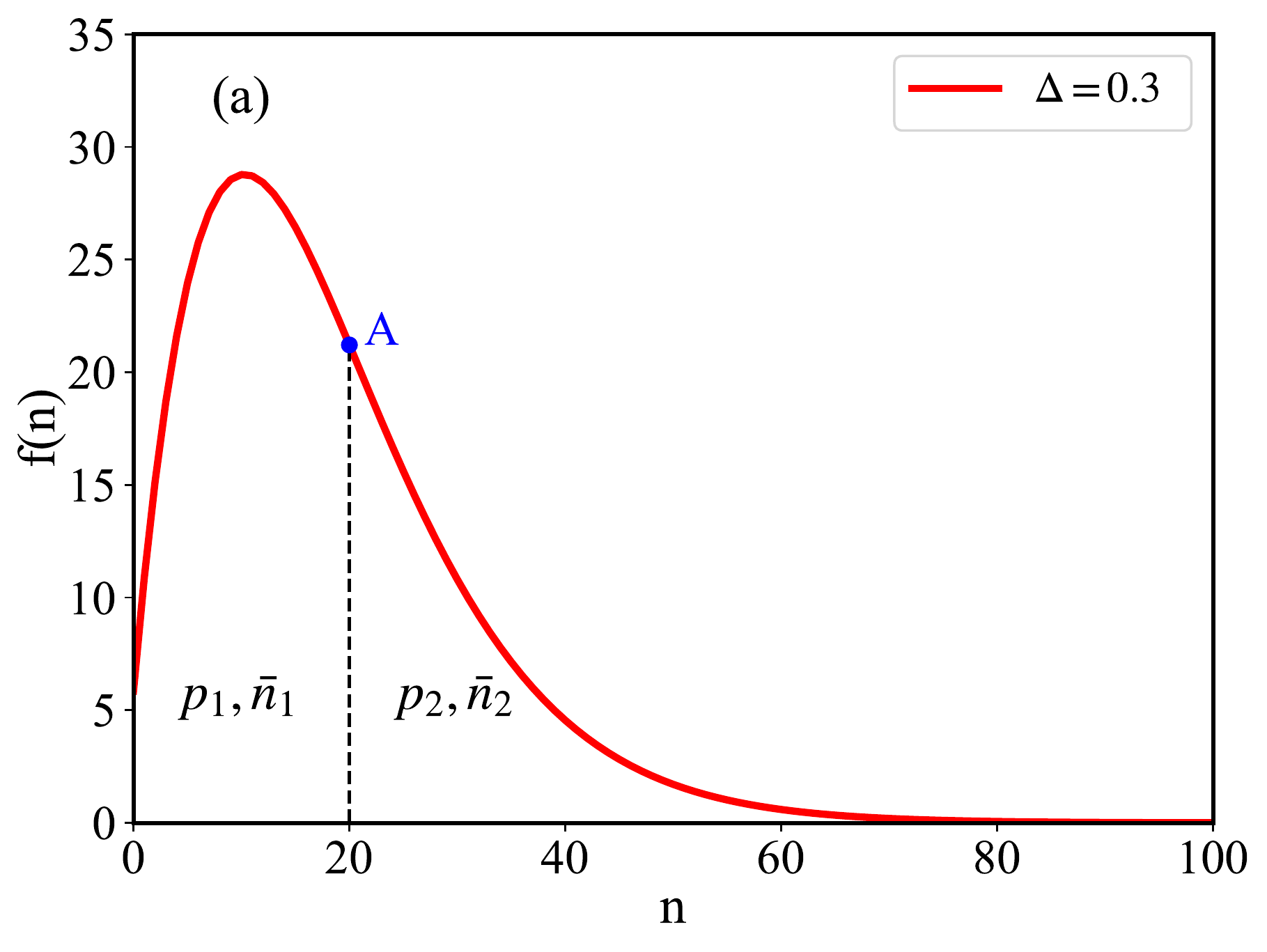}	
\includegraphics[width=8.8cm,height=6cm]{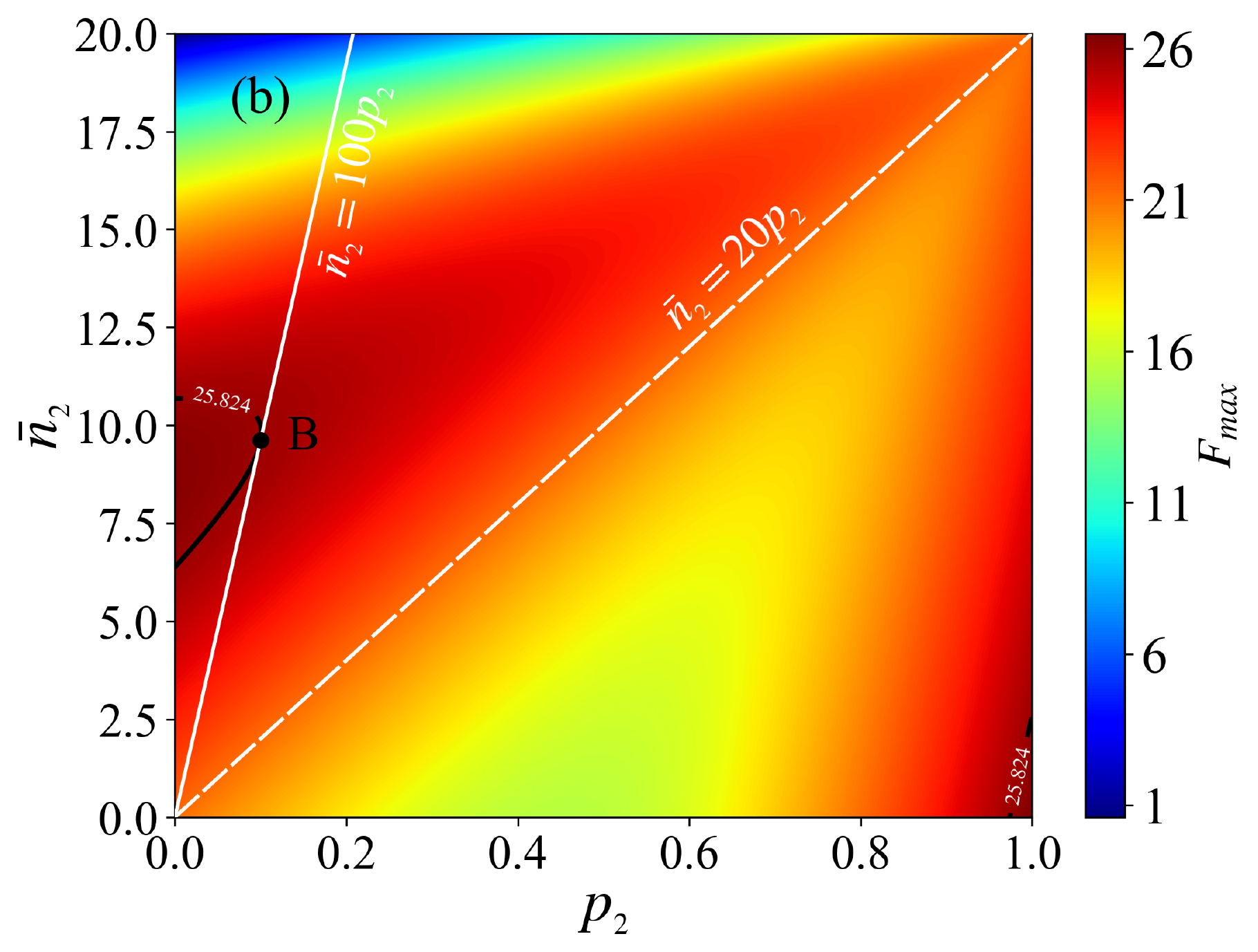}
\includegraphics[width=8.8cm,height=6cm]{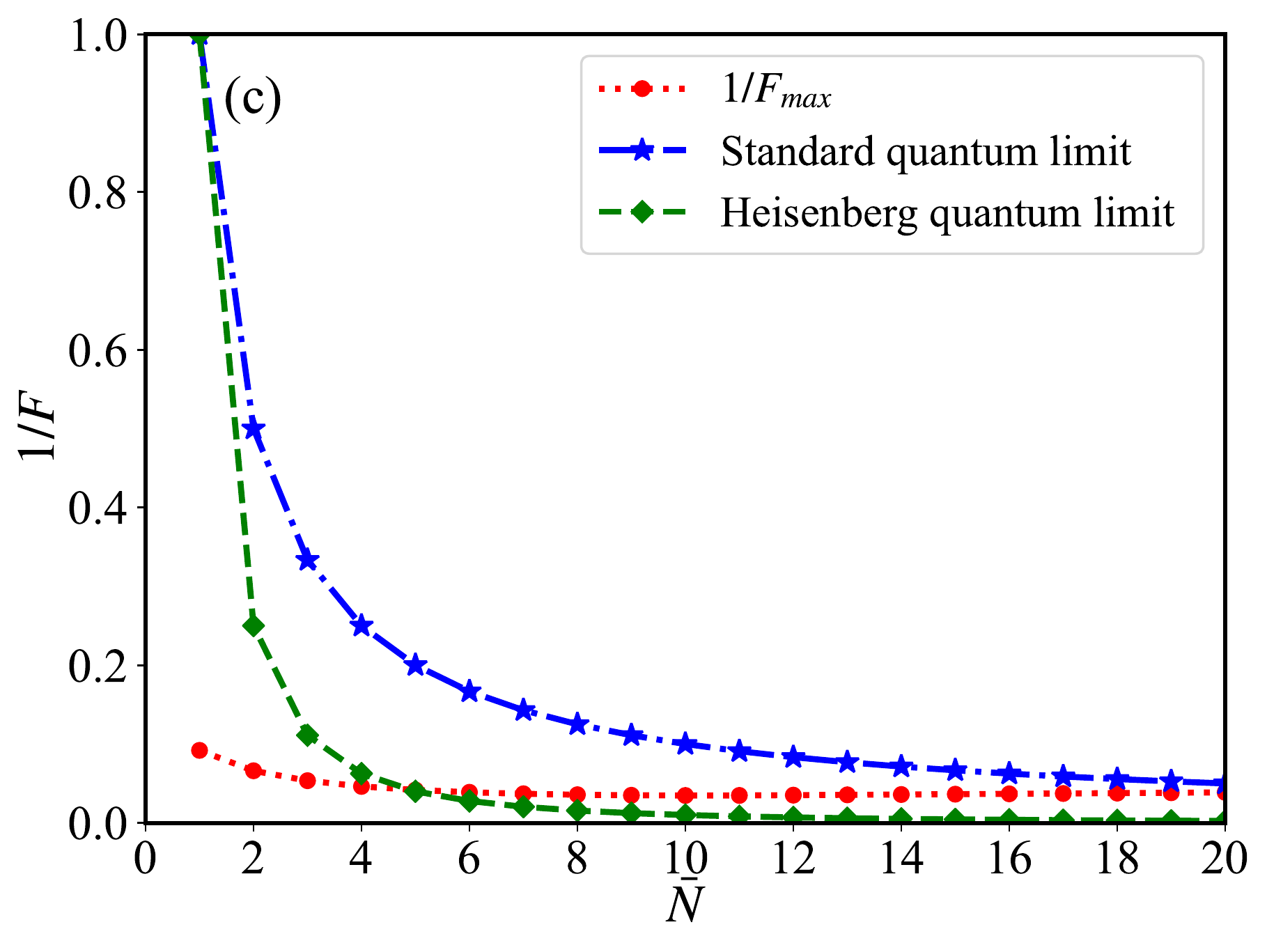}
\caption{\label{fig3}(a) Image of the variation of $f(n)$ with the number $n$ of photons. (b) Variation of the maximum quantum Fisher information $F_{max}$ with the probability distribution $p_{2}$ and the average photon number $\bar{n}_{2}$ in the second interval when the initial average photon number $\bar{N} = 20$. (c) Variation of the inverse of the maximum quantum Fisher information $F_{max}$ with different initial average photon numbers $\bar{N}$. The value of parameters are $v=1$, $\Delta=0.3$ \cite{PhysRevA.96.020301, PhysRevA.86.012107}.}
\end{figure}

In this section, our aim is to find the initial state of the optical field in the Landau-Zener-Jaynes-Cummings model when the quantum Fisher information of a parameter takes a maximum value for an equal initial average photon numbers. Quantum parameter estimation with the Landau-Zener transition has been studied in Ref.~\cite{PhysRevA.96.020301}, but quantum parameter estimation in photon-assisted Landau-Zener transition has not been studied yet. In Ref.~\cite{PhysRevA.86.012107}, the authors consider a Landau-Zener-Jaynes-Cummings model with the following Hamiltonian
\begin{equation}
\hat{H}=-\frac{vt}{2}\hat{\sigma}_{z}-\frac{\Delta}{2}(\hat{a}\hat{\sigma}_{+}+\hat{\sigma}_{-}\hat{a}^{\dagger}), \label{70}
\end{equation}
where $\hat{a}^{\dagger}$ and $\hat{a}$ respectively represent the creation and annihilation operators of bosons, $\hat{\sigma}_{\pm}=\hat{\sigma}_{x}\pm\hat{\sigma}_{y}$ and $\hat{\sigma}_{i} (i=x, y, z)$ is Pauli matrices of two-level system. $v$ is the speed of the sweep assumed to be positive and $\Delta$ is the level splitting at the transtion time $t=0$. Suppose the initial state of the system is
\begin{equation}
\left|\Phi(0)\right\rangle=\left|e\right\rangle\otimes\left|\psi(0)\right\rangle. \label{71}
\end{equation}
It is a tensor product of the excited state $\left|e\right\rangle$ of the two-level system with an arbitrary optical field state $\left|\psi(0)\right\rangle$. When $vt$ is switched from a large negative value to a large positive value, with the asymptotes of the parabolic cylinder functions, we can obtain the state of the total system as follows
\begin{equation}
\left|\Phi(t=\infty)\right\rangle=\sum_{n=0}^{\infty}C_{n}[A_{n}\left|e, n\right\rangle+B_{n}\left|g, n+1 \right\rangle], \label{72}
\end{equation}
where $C_{n}=\langle n\left|\psi(0)\right\rangle$ is the inner product between the number state and the initial state of the optical field, $A_{n}\approx\exp(-\pi\delta_{n})$, $\delta_{n}=\Delta^{2}(n+1)/(4v)$ and $\left|A_{n}\right|^{2}+\left|B_{n}\right|^{2}=1$. For the parameter $\Delta$, its corresponding quantum Fisher information on the pure state $\left|\Phi(t=\infty)\right\rangle$ is
\begin{equation}
F=4[\langle\partial_{\Delta}\Phi(\infty)|\partial_{\Delta}\Phi(\infty)\rangle-|\langle\Phi(\infty)|\partial_{\Delta}\Phi(\infty)\rangle|^{2}]. \label{73}
\end{equation}
Substituting Eq.~(\ref{72}) into Eq.~(\ref{73}) yields
\begin{equation}
F=\sum_{n=0}^{\infty}\left|C_{n}\right|^{2}f(n), \label{74}
\end{equation}
where 
\begin{equation}
f(n)=\frac{16\pi^{2}\delta_{n}^{2}e^{-2\pi\delta_{n}}}{\Delta^{2}(1-e^{-2\pi\delta_{n}})}. \label{75}
\end{equation}
Obviously, $F$ is an expected value of $f(n)$ on the probability distribution $\left|C_{n}\right|^{2}$.

The image of the function of $f(n)$ is shown in Fig.~\ref{fig3}(a). For the sake of discussion, we take $\Delta=0.3$, and we truncate the Hilbert space at the 100-photon Fock state $\left|n=100\right\rangle$. First, we assume that the independent variable $n$ is continuous, and then by the equation $d^{2}f(n)/dn^{2}=0$ we can get the transverse coordinate $n=20.83$ of the inflection point of $f(n)$. In fact, $n$ is a discrete integer, so we choose the transverse coordinate of the inflection point as $n=20$. That is, the blue point A in Fig.~\ref{fig3}(a) is the inflection point of $f(n)$. Obviously, the average slope of $f(n)$ in the interval $[0, 20)$ is gradually decreasing, while the average slope in the interval $[20, 100]$ is gradually increasing. Then, we assume that the initial mean photon number $\bar{N} = 20$ of the optical field. we assign the mean photon number $\bar{n}_{1}$ and probability $p_{1}$ in the interval $[0, 20)$ and the mean photon number $\bar{n}_{2}$ and probability $p_{2}$ in the interval $[20, 100]$. According to the theory we proved, the maximum quantum Fisher information is as follows
\begin{eqnarray}
F_{max}&=&(1-p_{2})f\Bigg(\frac{20-\bar{n}_{2}}{1-p_{2}}\Bigg)+\frac{100p_{2}-\bar{n}_{2}}{80}f(20) \nonumber \\
&&+\frac{\bar{n}_{2}-20p_{2}}{80}f(100), \label{76}
\end{eqnarray}
where $\bar{n}_{2}$ and $p_{2}$ satisfy these inequalities, $0\leq\bar{n}_{2} \leq 20$, $0\leq p_{2}\leq 1$, and $0\leq\frac{20-\bar{n}_{2}}{1-p_{2}}\leq20\leq\frac{\bar{n}_{2}}{p_{2}}\leq100$. The image of the binary function $F_{max}$ is shown in Fig.~\ref{fig3}(b). Here, the interval ranges of the independent variables $\bar{n}_{2}$ and $p_{2}$ satisfying the above inequality are located above the white dotted line and below the white solid line. The black solid line is the contour line when $F_{max}=25.824$. Obviously, the quantum Fisher information takes the maximum value at the point B where the black contour line and the white dashed line are tangent to each other.

Substituting the white solid line equation $\bar{n}_{2}=100p_{2}$ into Eq.~(\ref{76}), then in the interval $[0, 0.2]$ of $p_{2}$, the horizontal coordinate $p_{2}=0.0949$ and the vertical coordinate $\bar{n}_{2}=9.49$ of the point B can be obtained by equation $d F_{max}(p_{2})/dp_{2}=0$ and equation $\bar{n}_{2}=100p_{2}$. Also, since $p_{1}+p_{2}=1$ and $\bar{n}_{1}+\bar{n}_{2}=20$, we get $p_{1}=0.9051$ and $\bar{n}_{1}=10.51$. Considering that the average slope of $f(n)$ is getting smaller in the interval $[0, 20)$ and in the interval $[20, 100]$ is getting larger. Thus,  the probability distribution when F takes its maximum value is
\begin{eqnarray}
\left|C_{n}\right|^{2}&=&0.9051(0.388\delta_{11,n}+0.612\delta_{12,n})+0.0949\delta_{100,n} \nonumber \\
&\approx&0.3512\delta_{11,n}+0.5539\delta_{12,n}+0.0949\delta_{100,n}. \label{77}
\end{eqnarray}
Then, the optimal initial state of the optical field is
\begin{eqnarray}
\psi(0)=\sqrt{0.3512}\left|11\right\rangle+\sqrt{0.5539}\left|12\right\rangle+\sqrt{0.0949}\left|100\right\rangle. \label{78} \nonumber \\
\end{eqnarray}
The physical meaning of this result is clear. The maximum values of $f(n)$ are distributed around $f(11)$, so if the expected value of $f(n)$ is maximized, then most of the probability should be distributed around $n=11$. Then a small portion of the probability is distributed at the maximum value of $n=100$ such that the constraint $\bar{N}=20$ is satisfied. The maximum quantum Fisher information $F_{max}=25.824$ and its corresponding quantum Cramer-Rao bound is $(\varDelta\Delta)^{2}\geq \frac{1}{25.824}$, which beyond standard quantum limit ($\sim\frac{1}{20}$). In Fig.~\ref{fig3}(c), we plot the variation of the maximum quantum Fisher information with different initial average photon numbers. We find that when the initial average photon number $\bar{N}<5$, the corresponding quantum Cramer-Rao limit exceeds the Heisenberg quantum limit. When $5 \leq \bar{N} \leq 20$, the quantum Cramer-Rao limit of the system is able to exceed the standard quantum limit.

\subsection{The optimal state when the T-C quantum battery has the maximum stored energy and the average charging power}

\begin{figure}[t]
\centering
\includegraphics[width=8.8cm,height=6cm]{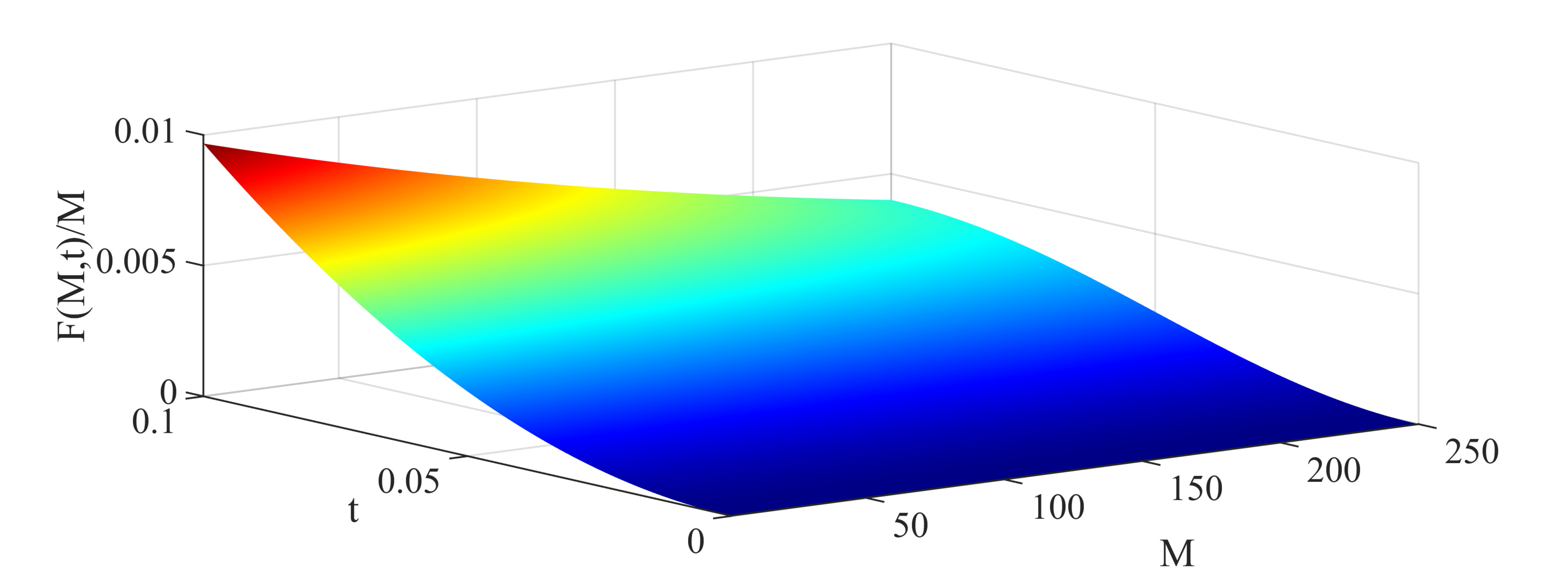}% Here is how to import EPS art
\caption{\label{fig4}Average slope $F(M,t)/M$ of $F(M,t)$ as a function of the time $t$ and the initial average photon number $M$.}
\end{figure}

We consider a quantum battery model, which consists of a cavity field and $N$ identical two-level atoms. The cavity field is the energy supply object, and the atomic ensemble is the energy storage object. When the photon-atom coupling strength is much smaller than the eigenfrequency of the cavity field or the transition frequency of the atom, and in a short time approximation, the interaction between the cavity field and $N$ identical two-level atoms is described by the Tavis-Cummings(TC) Hamiltonian. Therefore, this quantum battery is called TC quantum battery(TCQB) \cite{PhysRevA.104.043706}. Here we use our theorem to study the initial state of the cavity field that gives the TC quantum battery its maximum stored energy and maximum average charging power. We found that the stored energy $E(t)$ and the average charging power $P(t)$ of the TC quantum battery are the expected values of $F(M, t)$ and $F(M, t)/t$ under the probability distribution $p(n)$, respectively, where $p(n)$ is the probability distribution of the cavity initial state in the number states. The specific expressions for $E(t)$, $P(t)$, $F(M, t)$ are given in Ref.~\cite{PhysRevA.104.043706}. Here, we are only interested in the time when the stored energy and the average charging power of the TCQB first reach their maximum. The image of the function of $F(M, t)$ is shown in Fig.~\ref{fig4}. We can find that the average slope of the function $F(M,t)$ increases as $M$ increases at the same time $t$. According to the theory we proved, for an equal initial average photon number $\bar{n}$, the optimal initial states are number state $|\bar{n}\rangle$ and $\sqrt{1-(\bar{n}-[\bar{n}])}|[\bar{n}]\rangle+\sqrt{\bar{n}-[\bar{n}]}|[\bar{n}]+1\rangle$ when $\bar{n}$ is an integer and a non-integer, respectively. $[\bar{n}]$ represents the integer part of $\bar{n}$.

\section{\label{Sec:4} Conclusion }
Under the two constraints of a certain probability and a certain expected value of the independent variable $X$, We have proved four inequalities, and two of them can be reduced to Jensen's inequalities. The four inequalities can give the probability distribution when any monotonic function $F(X)$ takes the maximum or minimum expected value, and the maximum and minimum expected values of the monotonic function $F(X)$. Subsequently, we find that for a non-monotonic function $H(X)$, it can be divided into multiple monotonic intervals with respect to $X$. We find that we can transform finding the maximum and minimum expected value of $H(X)$ into the problem of finding the extreme value of a function $\overline{H(X)}_{max}$ with multiple variables. Finally, we present three applications of our proved maximum and minimum expected value theorems in quantum information processing. Our theory successfully explains why an arbitrary input path symmetric entangled state is superior to the NOON state in quantum parameter estimation. For an equal total input average photon numbers, we find the optimal input path symmetric state that makes the quantum Fisher information take the maximum value. When studying the quantum parameter estimation in Landau-Zener-Jaynes-Cummings model, we found the optimal initial state of the optical filed, which makes the system has the maximum quantum Fisher information. When studying the initial states that make the TC quantum battery has the maximum energy storage and the maximum average charging power, for an equal initial average photon numbers, we find that the optimal initial states are the number state and the superposition of the number state when the initial average photon number is an integer and a non-integer, respectively.

\begin{acknowledgments}
X.G.W was supported by the NSFC through Grant No.~11935012 and No.~11875231, the National Key Research and Development Program of China (Grants No.~2017YFA0304202 and No.~2017YFA0205700). W.J.L. was supported by the NSFC( Grants No. 11947069) and the Scientific Research Fund of Hunan Provincial Education Department( Grants No. 20C0495). 
\end{acknowledgments}  

\appendix

\section{\label{Appendix A}A RIGOROUS PROOF OF THE LEFT INEQUALITIES IN THE TWO THEOREMS}
Here, we give a specific proof of the two inequalities (\ref{8}) and (\ref{9}) when $x_{m}<\bar{n}/\bar{p}<x_{m+1}$. The difference of the expected value of $F(X)$ between the partical probability distribution $p_{1}(X)$ and an arbitrary probability distribution $p(X)$ is 
\begin{equation}
\varDelta F=\sum_{i=1}^{k}p_{1}(x_{i})F(x_{i})-\sum_{i=1}^{k}p(x_{i})F(x_{i}). \label{A1}
\end{equation}
Substituting Eq.~(\ref{6}) into the above equation yields
{\allowdisplaybreaks
\begin{eqnarray}
\varDelta F&=&\bar{p}F(x_{m})-\sum_{i=1}^{k}p(x_{i})F(x_{i})+\frac{\bar{n}-x_{m}\bar{p}}{x_{m+1}-x_{m}} \nonumber \\
&&\times [F(x_{m+1})-F(x_{m})] \nonumber \\
&=&\sum_{i=1}^{k}p(x_{i})[F(x_{m})-F(x_{i})]+\frac{\bar{n}-x_{m}\bar{p}}{x_{m+1}-x_{m}} \nonumber \\
&&\times [F(x_{m+1})-F(x_{m})] \nonumber \\
&=&\sum_{i=1}^{m-1}p(x_{i})[F(x_{m})-F(x_{i})]+\sum_{j=1}^{d}p(x_{m+j})[F(x_{m})  \nonumber \\
&&-F(x_{m+j})]+\frac{\bar{n}-x_{m}\bar{p}}{x_{m+1}-x_{m}} [F(x_{m+1})-F(x_{m})], \label{A2} \nonumber \\
\end{eqnarray}}
where $d=k-m$.
 
We rewrite Eq.~(\ref{4}) and Eq.~(\ref{5}) in the following forms
\begin{eqnarray}
&&\sum_{i=1}^{m-1}p(x_{i})+p(x_{m})+\sum_{j=1}^{d}p(x_{m+j})=\bar{p}, \label{A3} \\
&&\sum_{i=1}^{m-1}x_{i}p(x_{i})+x_{m}p(x_{m})+\sum_{j=1}^{d}x_{m+j}p(x_{m+j})=\bar{n}. \nonumber \\ \label{A4}
\end{eqnarray}
From the last two equations, we can obtain the following equation
\begin{equation}
\sum_{i=1}^{m-1}p(x_{i})(x_{m}-x_{i})+(\bar{n}-x_{m}\bar{p})=\sum_{j=1}^{d}p(x_{m+j})(x_{m+j}-x_{m}). \label{A5}
\end{equation}
We divide $p(x_{i})$, $(\bar{n}-x_{m}\bar{p})$ and $(x_{m+1}\bar{p}-\bar{n})/(x_{m+1}-x_{m})$ into $d$ parts, and thier $j$-th parts are respectively as follows
\begin{eqnarray}
&&p_{j}(x_{i})=\frac{(x_{m+j}-x_{m})p(x_{m+j})}{\sum_{k=1}^{d}p(x_{m+k})(x_{m+k}-x_{m})}p(x_{i}), \label{A6} \\
&&(\bar{n}-x_{m}\bar{p})_{j}=\frac{(x_{m+j}-x_{m})p(x_{m+j})(\bar{n}-x_{m}\bar{p})}{\sum_{k=1}^{d}p(x_{m+k})(x_{m+k}-x_{m})} , \label{A7}\\
&&\Bigg[\frac{x_{m+1}\bar{p}-\bar{n}}{x_{m+1}-x_{m}}\Bigg]_{j}=\frac{(x_{m+j}-x_{m})p(x_{m+j})}{\sum_{k=1}^{d}p(x_{m+k})(x_{m+k}-x_{m})} \nonumber \\
&&\hspace{1cm}\times\Bigg[\sum_{i=1}^{m}p(x_{i})-\frac{\bar{n}-x_{m}\bar{p}}{x_{m+1}-x_{m}}\Bigg]+p(x_{m+j}) .  \label{A8}
\end{eqnarray}
It is easy to verify that $p_{j}(x_{i})\geq0$, $(\bar{n}-x_{m}\bar{p})_{j}\geq0$,  $[(x_{m+1}\bar{p}-\bar{n})/(x_{m+1}-x_{m})]_{j}\geq0$, and $\sum_{j=1}^{d}p_{j}(x_{i})=p(x_{i})$, $\sum_{j=1}^{d}(\bar{n}-x_{m}\bar{p})_{j}=(\bar{n}-x_{m}\bar{p})$,    $\sum_{j=1}^{d}[(x_{m+1}\bar{p}-\bar{n})/(x_{m+1}-x_{m})]_{j}=[(x_{m+1}\bar{p}-\bar{n})/(x_{m+1}-x_{m})]$.
We can rewrite Eq.~(\ref{A5}) with the following form
\begin{eqnarray}
&&\sum_{j=1}^{d}\Big[\sum_{i=1}^{m-1}p_{j}(x_{i})(x_{m}-x_{i})+(\bar{n}-x_{m}\bar{p})_{j}\Big] \nonumber \\
&=&\sum_{j=1}^{d}p(x_{m+j})(x_{m+j}-x_{m}). \label{A9}
\end{eqnarray}
Substituting Eq.~(A6) and Eq.~(A7) into Eq.~(A9), we can obtain the following expression
\begin{equation}
p(x_{m+j})=\frac{1}{x_{m+j}-x_{m}}\Big[\sum_{i=1}^{m-1}p_{j}(x_{i})(x_{m}-x_{i})+(\bar{n}-x_{m}\bar{p})_{j}\Big] . \label{A10}
\end{equation} 
It is worth noting that we were able to obtain the last equation because of Eq.~(\ref{A6}) and Eq.~(\ref{A7}) hold. From Eq.~(\ref{A6}), Eq.~(\ref{A7}), and Eq.~(\ref{A8}), the following two equations hold
\begin{eqnarray}
&&\Bigg[\frac{\bar{n}-x_{m}\bar{p}}{x_{m+1}-x_{m}}\Bigg]_{j}+\Bigg[\frac{x_{m+1}\bar{p}-\bar{n}}{x_{m+1}-x_{m}}\Bigg]_{j} \nonumber\\
&&=\sum_{i=1}^{m}p_{j}(x_{i})+p(x_{m+j}), \label{A11}\\
&&x_{m}\Bigg[\frac{x_{m+1}\bar{p}-\bar{n}}{x_{m+1}-x_{m}}\Bigg]_{j}+x_{m+1}\Bigg[\frac{\bar{n}-x_{m}\bar{p}}{x_{m+1}-x_{m}}\Bigg]_{j}\nonumber \\
&&=\sum_{i=1}^{m}x_{i}p_{j}(x_{i}) +x_{m+j}p(x_{m+j}), \label{A12}
\end{eqnarray}
where $[(\bar{n}-x_{m}\bar{p})/(x_{m+1}-x_{m})]_{j}=(\bar{n}-x_{m}\bar{p})_{j}/(x_{m+1}-x_{m})$. Eq.~(\ref{A11}) indicates that after dividing the arbitrary probability distribution $p(X)$ and the special probability distribution $p_{1}(X)$ into d groups, their $j$-th group probability are equal. Eq.~(\ref{A12}) represents the equal expected value of $X$ under this equal probability splitting method. It is the reason that we call this method as the equal probability and equal expected value splitting method.
Substituting Eq.~(\ref{A10}) into Eq.~(\ref{A2}) yields
\begin{eqnarray}
\varDelta F&=&\sum_{j=1}^{d}\sum_{i=1}^{m-1}p_{j}(x_{i})(x_{m}-x_{i})\Bigg[\frac{F(x_{m})-F(x_{i})}{x_{m}-x_{i}}-\nonumber \\
&&\frac{F(x_{m+j})-F(x_{m})}{x_{m+j}-x_{m}}\Bigg]+\sum_{j=2}^{d}(\bar{n}-x_{m}\bar{p})_{j} \nonumber \\
&&\times\Bigg[\frac{F(x_{m+1})-F(x_{m})}{x_{m+1}-x_{m}}-\frac{F(x_{m+j})-F(x_{m})}{x_{m+j}-x_{m}}\Bigg]. \label{A13} \nonumber \\
\end{eqnarray}
If $x_{m}=\bar{n}/\bar{p}$, then, Eq.~(\ref{A13}) can be reduced to Eq.~(\ref{28}). Since $p_{j}\geq0$, $x_{m}-x_{i}\geq0$ and $(\bar{n}-x_{m}\bar{p})_{j}\geq0$, if the average slope of $F(X)$ is getting smaller, then, $\varDelta F\geq0$, i.e., 
\begin{eqnarray}
\frac{x_{m+1}\bar{p}-\bar{n}}{x_{m+1}-x_{m}}F(x_{m})&&+\frac{\bar{n}-x_{m}\bar{p}}{x_{m+1}-x_{m}}F(x_{m+1})\nonumber \\
&&\geq\sum_{i=1}^{k}p(x_{i})F(x_{i}). \label{A14}
\end{eqnarray}
Conversely, if the average slope of $F(X)$ is getting larger, then, $\varDelta F\leq0$, i.e.,
\begin{eqnarray}
\frac{x_{m+1}\bar{p}-\bar{n}}{x_{m+1}-x_{m}}F(x_{m})&&+\frac{\bar{n}-x_{m}\bar{p}}{x_{m+1}-x_{m}}F(x_{m+1})\nonumber \\
&&\leq\sum_{i=1}^{k}p(x_{i})F(x_{i}). \label{A15}
\end{eqnarray}

\section{\label{Appendix B}DERIVATION OF EQ.~(\ref{34}) }
Here, we give a derivation of Eq.~(\ref{34})
{\allowdisplaybreaks
\begin{eqnarray}
\varDelta F
&=&\frac{\sum_{i=1}^{k}p(x_{i})(x_{k}-x_{i})}{x_{k}-x_{1}}F(x_{1})-\sum_{i=1}^{k}p(x_{i})F(x_{i})\nonumber \\
&&+\frac{\sum_{i=1}^{k}p(x_{i})(x_{i}-x_{1})}{x_{k}-x_{1}}F(x_{k})\nonumber \\
&=&p(x_{1})F(x_{1})+\frac{\sum_{i=2}^{k-1}p(x_{i})(x_{k}-x_{i})}{x_{k}-x_{1}}F(x_{1}) \nonumber \\
&&+p(x_{k})F(x_{k})+\frac{\sum_{i=2}^{k-1}p(x_{i})(x_{i}-x_{1})}{x_{k}-x_{1}}F(x_{k}) \nonumber \\
&&-p(x_{1})F(x_{1})-p(x_{k})F(x_{k})-\sum_{i=2}^{k-1}p(x_{i})F(x_{i}) \nonumber \\
&=&\sum_{i=2}^{k-1}\frac{p(x_{i})(x_{i}-x_{1})(x_{k}-x_{i})}{x_{k}-x_{1}}\Bigg[\frac{F(x_{k})-F(x_{i})}{x_{k}-x_{i}} \nonumber  \\
&&-\frac{F(x_{i})-F(x_{1})}{x_{i}-x_{1}}\Bigg]. \label{B1}
\end{eqnarray}
}

\section{\label{Appendix C}DERIVATION OF EQ.~(\ref{66}) }
Here, we give a derivation of Eq.~(\ref{66})
{\allowdisplaybreaks
\begin{eqnarray}
&&\varDelta F_\text{NOON}=\sum_{n=0}^{m}p(n)\Big[\bar{n}^{2}-\frac{n^{2}}{1+p(0)}\Big]+\sum_{j=1}^{d}p(m+j)\nonumber \\
&&\hspace{2cm}\times\Big[\bar{n}^{2} -\frac{(m+j)^{2}}{1+p(0)}\Big] \nonumber \\
&&\hspace{1.5cm}=\sum_{n=0}^{m}\sum_{j=1}^{d}p^{j}(n)\Bigg[\Big[\bar{n}^{2}-\frac{n^{2}}{1+p(0)}\Big]+\Big[\bar{n}^{2}-\nonumber \\
&&\hspace{2cm}\frac{(m+j)^{2}}{1+p(0)}\Big]\frac{\bar{n}-\frac{n}{1+p(0)}}{\frac{m+j}{1+p(0)}-\bar{n}} \Bigg] \nonumber\\
&&\hspace{1.5cm}=\sum_{n=0}^{m}\sum_{j=1}^{d}p^{j}(n)\Big[\bar{n}-\frac{n}{1+p(0)}\Big]\times \nonumber \\
&&\Bigg[\frac{(1+p(0))\bar{n}^{2}-n^{2}}{(1+p(0))\bar{n}-n}-\frac{(m+j)^{2}-\bar{n}^{2}(1+p(0))}{m+j-\bar{n}(1+p(0))}\Bigg]. \label{C1}
\end{eqnarray}}
\nocite{*}

\bibliography{MaxMinExpectation}% Produces the bibliography via BibTeX.

\end{document}